# Modeling Genetic Networks from Clonal Analysis

Radhakrishnan Nagarajan

*Center on Aging, University of Arkansas for Medical Sciences*
*Address: Center on Aging, UAMS*
*629 Jack Stephens Drive, Room: 3105*
*Little Rock, AR 72205, USA*
*email: nagarajanradhakrish@uams.edu*

Jane E. Aubin

*Department of Medical Genetics and Microbiology*
*University of Toronto*
*Room 6230 Medical Sciences Building*
*1 King's College Circle*
*Toronto, Ontario M5S 1A8, Canada*
*email: jane.aubin@utoronto.ca*

Charlotte A. Peterson

*Center on Aging, University of Arkansas for Medical Sciences*
*Central Arkansas Veterans Health Care System*
*Address: Center on Aging, UAMS*
*629 Jack Stephens Drive, Room: 3121*
*Little Rock, AR 72205, USA*
*email: petersoncharlottea@uams.edu*

*Corresponding Author:*

Radhakrishnan Nagarajan
*Center on Aging*
*University of Arkansas for Medical Sciences*
*629 Jack Stephens Drive, Room: 3105*
*Little Rock, AR 72205, USA*
*Phone: (501) 526-7461*
*Fax: (501) 526-5830*
*email: nagarajanradhakrish@uams.edu*

**Abstract**

In this report a systematic approach is used to determine the approximate genetic network and robust dependencies underlying differentiation. The data considered is in the form of a binary matrix and represent the expression of the nine genes across the ninety-nine colonies. The report is divided into two parts: the first part identifies significant pair-wise dependencies from the given binary matrix using linear correlation and mutual information. A new method is proposed to determine statistically significant dependencies estimated using the mutual information measure. In the second, a Bayesian approach is used to obtain an approximate description (equivalence class) of network structures. The robustness of linear correlation, mutual information and the equivalence class of networks is investigated with perturbation and decreasing colony number. Perturbation of the data was achieved by generating bootstrap realizations. The results are refined with biological knowledge. It was found that certain dependencies in the network are immune to perturbation and decreasing colony number and may represent robust features, inherent in the differentiation program of osteoblast progenitor cells. The methods to be discussed are generic in nature and not restricted to the experimental paradigm addressed in this study.

## 1. Introduction

Biological processes such as cell differentiation are mediated by specific networks of genes, which interact with one another. Such systems evolved with time and can be aptly characterized as coupled nonlinear dynamical systems. Dynamical systems can be broadly classified into linear and nonlinear systems. While the former can exhibit interesting behavior only in the presence of noise, the latter can give rise to a wide range of behavior even in the absence of noise. Current techniques used in modeling genetic networks rely on the concepts of biochemical networks (Goldbeter, 1996; Jacob and Monod, 1961; Kauffman, 1969; Tyson et al, 2001; Yagil and Yagil, 1971). The interaction between genes is often modeled as the outcome of a deterministic sequence of events (Gardner et al., 2000; Huang and Ferrell, 1996). However, recent reports have indicated that, in spite of the deterministic phenotypic outcome, the underlying dynamics may be noisy (Hasty et al, 2000; McAdams and Arkin, 1997; McAdams and Arkin, 1999). Noise can be categorized broadly into either dynamical or measurement noise. While the former is coupled to the dynamics, the latter occurs externally (Fig. 1). Various factors that contribute to dynamical noise include: fluctuations in protein concentrations, variation in cell-to-cell switching time and micro-environment. The phenotypic outcome is deterministic in the average sense and has been attributed to population transcriptional cooperation, checkpoints, and redundancies in genes and regulatory pathways, (McAdams and Arkin, 1999). In a recent study, it was demonstrated how noise can be used to control switching and regulation of gene expression (Hasty et al., 2000). On the other hand, a purely nonlinear deterministic approach has been used to model gene interactions (Gardner et al., 2000). The gene expression values are determined with a measurement device, which acts as a black box between the true biological activity and the output (Fig. 1). Thus, what is externally observed is the output of the measurement device and not the true biological phenomena. It might not be surprising to note that the input and output of the measurement device may be nonlinearly related, corrupted with measurement noise. Thus the choice of the mathematical technique in explaining the observed phenomenon is an open question. Given the above complexities, a pragmatic approach would be to use a combination of linear and nonlinear measures (see Sec. 2).

There have been a number of reports on modeling genetic networks as Boolean networks (Glass and Hill, 1998; Kauffman, 1993), where each gene can be either in an ON or an OFF state. The switching of a gene is commonly represented by a sigmoidal curve, whose steepness depends on the extent of transcriptional *cooperativity* (Goldbeter and Dupont, 1990; Huang and Ferrel, 1996). Temporal expression profile studies have also been used successfully in determining the network structure from such binary representations (Liang et al., 1998; Shmulevich et al., 2002). However, in the present report we address the problem of determining the network structure from gene expression patterns across replicate cell colonies, by observing a snap shot of the biological process represented by a binary matrix, i.e. each gene is either ON or OFF in the given time window across the replicate colonies.

*Clonal Analysis of Osteoblast Progenitor Cells*

The formation of a particular differentiated phenotype from a stem cell/progenitor is believed to follow a hierarchical sequence of events. Recent studies (Aubin and Triffit, 2001; Liu et al, 2003; Madras et al., 2002) have indicated that the osteoblast developmental program may be heterogeneous and probabilistic in nature. Madras et al., investigated the expression of nine osteoblast-associated genes in a window of observation. The genes consisted of: COLL (collagen I), ALP (alkaline phospotase), BSP (bone sialoprotein), OCN (osteocalcin), FGFR1 (fibroblast-growth factor receptor 1), PDGFR$\alpha$? (platelet-derived growth factor receptor $\alpha$), PTH1R (parathroid hormone/parathyroid hormone related protein receptor), PTHrP (parathyroid hormone related protein) OPN (osteopontin) across ninety-nine colonies cultured under identical conditions. These genes are asynchronously expressed during the course of osteoblast differentiation, however, there are some discrepancies in the literature regarding their sequence of expression during the osteoblast development [Madras et al., 2002]. The colony forming units osteoblasts (CFU-Os) were plated from rat-calvariae cells (21 day Wistar rat fetuses). While some genes are constitutively expressed (i.e. ON), others occur with a *finite probability* in a given colony (i.e. ON or OFF), (Aubin, 1998; Aubin et al., 1999; Madras et al., 2002). The ON/OFF of a gene was determined by quantifying the signal over background from digitized images of Southern blots by Phosphorimager. The normalized values represent the ratio of *gene A's* signal over total cDNA (used as housekeeping gene control in global amplification) signal for each gene in each sample, thereby generating the binary matrix, considered in this study. While there are technical issues with quantifying densitometry signals, these were controlled rigorously. In the analysis, any normalized signal of any magnitude with a specific gene probe that was over background was considered "ON". A statistical approach was subsequently used to determine pair-wise dependence and the temporal sequence of expression of the nine genes obtained across the ninety-nine colonies. The authors concluded that the probabilistic nature of gene expression could result in multiple pathways giving rise to the same phenotypic outcome. Determining dependencies and causal relationships between the genes from a snap shot of the biological process can be quite challenging. In this report, a systematic approach different from that discussed in (Madras et al, 2002) is used to infer the dependencies between the genes and the approximate network structure. The techniques are generic in nature and hence not restricted to the osteoblast differentiation data.

The report is organized as follows: in Section 2, the properties of entropy, linear correlation and mutual information are discussed. Entropy is a measure of uncertainty associated with expression of a gene, whereas correlation and mutual information measure the dependency between pairs of genes. The Fisher's exact test which is widely used in categorical data analysis is also used to capture non-random association between the genes. While the statistical significance of linear correlation and Fisher's exact test is well defined, that of mutual information is unclear. A resampling approach is proposed in Section 2, and used subsequently to determine statistically significant dependencies from mutual information estimates. In Section 3, the effect of perturbations and decreasing colony number on the entropy, linear correlation, Fisher's exact test and the mutual information estimates is investigated. Perturbations of the data were generated by bootstrap realizations. In Section 4, a Bayesian approach is used as a tool to determine the approximate network structure. Constraint-based and search algorithms are used to determine the equivalent class of networks. A brief summary of the results including pruning the network structure with biological knowledge and possible extensions is included under Section 5.

## 2. Pair-wise Dependencies and their Statistical Significance

*2.1 Pair-wise dependencies - Linear Correlation, Mutual Information and Fisher's Exact test*

The two measures commonly used to determine the dependencies between any two random variables ($x$ and $y$) are the correlation ($r_{xy}$) and the mutual information ($I_{xy}$), given by

$$r_{xy} = \frac{1}{N-1}\sum_{i=1}^{N} \frac{(x-m_x)(y-m_y)}{s_x s_y} \text{ where } |r_{xy}| \leq 1,\ r_{xy} = r_{yx} \text{ ..................(1)}$$

The self-information (entropy) is given by $H(x) = -\sum_x p_x \log_2 p_x$ Shannon (1948).

$$I_{xy} = \sum_{x,y} p_{xy} \log(\frac{p_{xy}}{p_x p_y}) = H(x) + H(y) - H(x,y) \text{ where } I_{xy} \geq 0,\ I_{xy} = I_{yx} \text{ ......(2)}$$

While $r_{xy}$ is a measure of linear dependence, $I_{xy}$ measures the generalized dependence and has been used in inferring genetic networks (Liang et al., 1998) and nonlinear studies (Fraser and Swinney, 1986). From equations (1) and (2), the measures $I_{xy}$, $r_{xy}$ are symmetric and hence fail to indicate the *direction* of dependence. Pair-wise dependencies of a network composed of several genes, can be represented as a symmetric matrix. The diagonal elements of the correlation matrix, represents the auto-correlation $r_{xx}$ and is unity, whereas those of the mutual information matrix $I_{xx}$, represents the self-information or the entropy of the gene. The entropy of a gene across the independent cell colonies is a measure of the uncertainty in its expression. In the present experimental paradigm, the gene OPN was expressed in all the colonies and hence had an entropy zero. The mutual information $I_{xy}$ is a measure of the reduction in uncertainty of $x$ having observed $y$. For deterministic $y$ (ex: OPN) one fails to observe any reduction in uncertainty of $x$ i.e. $H(y) = 0$ implies $H(x,y) = H(y)$, resulting in $I_{xy} = 0$ and $r_{xy} = 0$. Hence we shall leave OPN out from subsequent discussions. An analytical relation relates $I_{xy}$ and $r_{xy}$ for normally distributed processes, given by $I_{xy} = -0.5\log(1 - r_{xy}^2)$. The Fisher's exact test is ideal for determining non-random associations between categorical variables and has been used successfully by (Madras et al., 2002) to capture statistically significant dependencies between the genes.

### *2.2 Significance of the pair-wise dependencies*

#### *2.2.1 Resampling Techniques*

Resampling methods have been used as effective tools to assess statistical significance (Efron et al., 1994). In traditional tests, the statistical measure obtained on a sample, drawn from a given population, is compared to a *theoretical distribution* to assess its significance. However in resampling statistics, one generates an *empirical distribution* by computing the statistical measure on samples generated from the *same* empirical sample, hence the term resampling. Unlike traditional approaches, resampling implicitly assumes the given empirical sample to be a representative of the population, which is considered a drawback of this technique. While resampling without replacement is suited for correlated data (Schreiber, 1996), resampling with replacement is ideal for uncorrelated data. In the present study each colony is independent of the other and hence the choice of resampling with replacement (bootstrap) is justified. In a recent study, bootstrap realizations have been identified with perturbing the given data (Friedman et al., 2002).

*2.2.2 Statistical significance of the measures*

The p-value of the linear correlation and Fisher's exact test can be estimated from parametric distributions, however this is not the case for mutual information. An ad-hoc approach would be to choose a fixed threshold (θ) and assume all pair-wise dependencies whose mutual information is greater than (θ) to be significant. In a recent report, Butte and Kohane (2000) showed that the choice of the threshold (θ) could significantly alter the network architecture. Therefore, it is important to develop a statistically rigorous approach to avoid spurious conclusions. In this report we propose a method to determine statistically significant mutual information $I_{xy}$ estimates between a pair of genes ($x$ and $y$). A step-wise procedure is described below:

- Randomly choose $k$ colonies with replacement (bootstrap sample) and pair of genes $(x, y)$. The null hypothesis addressed here is that the expression of $(x, y)$ across the $k$ colonies is uncorrelated.
- Determine the mutual information between this pair $I_{xy}$.
- Randomly shuffle the ON/OFF states for these genes across the $k$ colonies. These shall be referred to as a *surrogate*. Compute the mutual information between the shuffled pair $I_{x,y}^{shuff}$. Generate 19 such surrogates. This corresponds to $a$ = 1/(19+1) = 0.05 (Hope, 1968; Theiler and Prichard, 1995).

- The $p$-value is given by

$$p = \frac{[\text{number of cases where } I_{xy} < I_{x,y}^{shuff} + 1]}{20}$$

  The numerator has a minimum value of one, which corresponds to the original pair.

- If $I_{xy} > I_{x,y}^{shuff}$ for each of the 19 surrogates then the null is rejected at $a = 0.05$.

- Repeat the above procedure for $N_B$ times. The power of the test is the fraction of the number of successful rejections of the null divided by $N_B$.

It should be noted that the random-shuffled surrogates retain the self-information, *H(x)* and *H(y)* of the genes, whereas the joint-entropy *H(x,y)* is destroyed; such surrogates have been classified under a special class called *constrained randomized surrogates* (Theiler and Prichard, 1995; Schreiber and Schmitz, 1996). Constrained realizations as opposed to typical realizations retain certain statistical features of the original data, imposed as constraints. The constraint here is on retaining the self-information. In the above algorithm, we determined the p-value corresponding to significance level $a$. This represents the Type I error, which is the odds of falsely rejecting a given null hypothesis, i.e. concluding two genes are pair-wise dependent, when they are not. The power of the statistical test is related to the Type II error ($b$) as (1-$b$) and represents rejecting the null when it is false. The power is determined by generating several independent bootstrap realizations, shown in the last step.

*2.2.3 Comment on Bonferroni correction:*

The Bonferroni correction has been used extensively in gene expression analyses (Wayne and McIntyre, 2002). However, there are some subtle details associated with the Bonferroni correction (Perneger, 1998), which brings up the question of whether it is appropriate for the analysis of gene expression data. The objective is to obtain an adjusted value ($a^{*} = a/n$) of the Type I error, for *each* of the $n$ individual tests so as to maintain an experiment-wise error rate of $a$. The null hypothesis addressed by the Bonferroni test is that the two groups of $n$ tests are not statistically different. In the present study, we have $n = 8$ genes in the network (excluding OPN, see Sec. 1), resulting in 28 pairs. A $p$-value lesser than $a^{*} = 0.05/28 = 0.0018$, obtained on *any one* of the pairs, rejects the null that *all* the 28 pairs are independent. It is important to evaluate the usefulness of the result in the present context. The phenotypic outcome in a given experimental paradigm is mediated by specific signaling pathways with interacting components and are therefore likely to be dependent. Thus, rejecting the null that the components are independent of one another seems to be of little use. The expression for the Bonferroni correction is obtained under the assumption that the pair-wise correlations are independent of one another, which is not true in general. This can lead to spurious conclusions, more importantly an increase in the Type II error and a reduction in statistical power. The impact of the Bonferroni correction in determining the dependencies with the correlation measure shall be discussed in the following section.

*Summary (Sec. 2): The choice of resampling approach as opposed to ad-hoc thresholding can be used to determine statistically significant pair-wise dependencies estimated by the mutual information measure.*

**3. Effect of number of colonies on pair-wise dependencies and entropy**

*3.1 Self-information (entropy) of the genes*

The entropy of a gene is a measure of the uncertainty or randomness associated with its expression. If there were only a single pathway to explain a given phenotypic outcome, then one would observe the same set of genes to be constitutively activated in each colony. However, the present experimental paradigm failed to exhibit such a behavior other than for OPN. The value of the entropy of binary sequence in $\log_2$ scale is bounded by [0, 1]. A value of zero is indicative of *determinism*, whereas a value of one is indicative of maximum *randomness*. The former can occur when a particular gene is constitutively ON or OFF across *all* the colonies; the latter is observed when the odds of a gene being present or absent in a given colony is half. Entropy of the eight genes, estimated by using 100 bootstrap realizations for sample size of 99, 66, and 33 colonies, is shown in Figure 2. The ranked entropy profile of the genes [*ALP, BSP, OCN, COLL, PDGFRa, FGFR1, PTHrP, PTH1R*] remained unchanged with perturbation and reduction in number of colonies.

*Summary (Sec. 3.1): Non-zero entropy of genes across replicate colonies indicates possible non-determinisitic nature of gene expression across the colonies. Thus it might be possible to achieve the end phenotype through multiple gene activation programs. The*

*ranked entropy profile of the genes is robust to bootstrapping (perturbation of the data) and decreasing number of colonies.*

While entropy is a useful measure of self-information of *a gene*, it does not provide insight into the dependencies *between genes*. Interestingly, we found the dependencies between the high-entropy genes to be robust to decreasing colony number and perturbations, discussed in the following section.

*3.2 Estimating pair-wise dependencies using Linear Correlation*

Pair-wise dependencies were estimated by sampling 99, 66 and 33 colonies from the given 99 colonies (Madras et al., 2002). One hundred bootstrap realizations were generated from the given data (99 colonies) for each sample size: 99, 66 and 33. This was done to determine the robustness of certain dependencies with perturbation of the given data and decreasing colony number. The mean and the variance of the correlation $(r)$ estimated for these sample sizes (99, 66 and 33 colonies) are shown in Figures 3b, 4b and 5b, respectively. The correlation obtained on the random shuffled surrogates, (Sec. 2.2.2) representing the *noise floor* is also shown as a reference. The power ($N_B$ = 100) estimated with Bonferroni correction ($a$ = 0.0018) and without Bonferroni correction ($a$ = 0.05).

*Without Bonferroni Correction*: The pair-wise dependencies that came through the power analysis with (power ~ 0.8), Figure 6a, were *(COLL–ALP), (COLL–OCN), (COLL-FGFR1) (ALP–BSP), (ALP–OCN), (ALP-PTH1R), (BSP–OCN)* and *(OCN–PTH1R)* by resampling 99 and 66 colonies. However, resampling with 33 colonies retained only four of the above pairs (power ~ 0.8) namely: *(COLL–ALP), (ALP–BSP), (ALP–OCN),* and *(BSP–OCN).* The number of dependencies obtained for the various colony sizes ($n_c$ = 11, 22, 33, 66 and 99) is enclosed under Table I. It should be noted that the number of dependencies (α = 0.05, power ~ 0.8) vary considerably with varying colony size.

*With Bonferroni Correction*: The pair-wise dependencies that came through (power ~ 0.8) by resampling 99 colonies were *(COLL–ALP), (COLL–OCN), (ALP–BSP), (ALP–OCN), (BSP–OCN)* and *(OCN–PTH1R.* Those obtained by resampling 66 were *(COLL–ALP), (ALP–BSP), (ALP–OCN), (BSP–OCN)* and those obtained by resampling 33 colonies was *(ALP–OCN).* These results confirm our earlier remark on the effect of the Bonferroni correction in reducing statistical power, Table I. As observed by Madras et al. (2002) the genes *PDGFRa* and *PTHrP* were independent of all the other genes and from one another. This is supported by the high p-value and low power obtained for these pairs.

*Summary (Sec. 3.2): The pair-wise dependencies estimated using linear correlation (a = 0.05, power ~ 0.8), that were robust to decreasing sample size ($n_c$ = 99, 66 and 33) and perturbation consisted of: (COLL–ALP), (ALP–BSP), (ALP–OCN), and (BSP–OCN)*

### *3.3 Estimating pair-wise dependencies using Fisher's exact test*

The Fisher's exact test had been used recently (Madras et al., 2002) to determine statistically significant dependencies in the given experimental paradigm. As opposed to linear correlation and mutual information, this test assumes that the given data lends itself into *categories*. Consider the binary expression of two genes $g_1$ and $g_2$ across n colonies. There are four possible occurrences namely ($g_1$, $g_2$) = (00, 01, 10 and 11). The number of occurrences of each of these combinations is represented by a 2 x 2 contingency table. The p-values are determined from hypergeometric distribution. Statistically significant genes ($a$ = 0.05) with (power ~ 0.8), Figure 6b, for varying colony sizes ($n_c$ = 11, 22, 33, 66 and 99) is enclosed under Table I. The statistically genes with ($a$ = 0.05, power ~ 0.8) were similar to those obtained using linear correlation. As with linear correlation the number of statistically significant genes vary considerably with varying colony numbers. This implies that there might not be any added advantage in using linear correlation over Fisher's exact test.

*Summary (Sec. 3.3): The pair-wise dependencies estimated using Fisher's exact test ($a$ = 0.05, power ~ 0.8), that were robust to decreasing sample size ($n_c$ = 99, 66 and 33) and perturbation consisted of: (COLL–ALP), (ALP–BSP), (ALP–OCN), and (BSP–OCN)*

### *3.4 Estimating pair-wise dependencies using Mutual Information*

The mean and the variance of the mutual information (Sec. 2.2.2) estimated from 100 bootstrap realizations with sample sizes: 99, 66 and 33 colonies are shown in Figures 3a, 4a and 5a, respectively. The noise floor is represented by value estimated on the surrogates (Sec. 2.2.2). Mutual information estimated from the correlation measure under normality assumption, as $I_{xy} = -0.5 \log(1 - r_{xy}^2)$ is also shown as reference. For the data considered in this study, the mutual information estimated using the normal approximation closely follows obtained using (2). However, this might not be true in general. The pair-wise dependencies were significant (a = 0.05 and power ~ 0.8), Figure 6c, across samples sizes (nc = 33, 66 and 99) were *(COLL–ALP), (COLL–OCN), (ALP–BSP), (ALP–OCN), (BSP–OCN) and (OCN–PTH1R).* Unlike linear correlation measure and Fisher's exact test, the number of statistically significant mutual information estimates ($a$ = 0.05, power ~ 0.8) do not change dramatically with decreasing number of colonies ($n_c$ = 33, 66 and 99). Thus mutual information is comparatively robust in capturing the pair-wise dependencies with decreasing colony number and perturbations. As with the correlation measure and Fisher's exact test, the genes *PDGFRa* and *PTHrP* were independent of all the other genes and from one-another and reflected by the low power and high p-value. This result is in agreement with those reported by (Madras et al., 2002).

*Summary (Sec. 3.4): The pair-wise dependencies estimated using mutual information ($a$ = 0.05, power ~ 0.8), that were robust to decreasing sample size ($n_c$ = 99, 66 and 33) and perturbation consisted of: (COLL–ALP), (COLL–OCN), (ALP–BSP), (ALP–OCN), (BSP–OCN) and (OCN-PTH1R)*

Although the genes ALP, BSP, OCN and COLL form the set of high entropy genes, the pair-wise dependencies COLL – ALP; ALP – OCN; ALP – BSP; BSP – OCN persist with decreasing sample sizes and perturbations. These were verified using correlation, Fisher's exact test and mutual information. Hence these might form the set of robust dependencies in the osteoblast differentiation program. While the colony numbers (nc =

22 and 11) are included in Table I, the number of dependencies dramatically reduce for these colony numbers, which implies that it might not be possible to statistically infer the dependencies for colony numbers lesser than ($n_c$ = 33).

The above analysis treats pair-wise dependencies as *mutually exclusive* events. However, when we connect these as a part of the network, it is possible that the dependencies may cease to exist. This is attributed to the fact that the pair-wise dependency may be *conditionally dependent* on the other elements in the network. The symmetric nature of the pair-wise dependency measures discussed above does not provide information regarding the directionality or more precisely the *causality* information of the network. In the following section, we use the concepts of Bayesian networks (Pearl, 2000) to infer the network structure from the given data, under certain generic conditions.

## 4. Bayesian Networks Approach to determine network structure

Bayesian networks (Pearl, 2000; Spirtes et al, 2001) have been used extensively to model conditional dependencies in a given data. Recent reports (Friedman et al., 2000) have discussed their application for analyzing gene expression data obtained from microarrays. In this report, Bayesian networks is used as a tool to determine the network structure from the given data. A brief introduction along with definitions and simple examples is provided for completeness. A rigorous theoretical treatment can be found elsewhere (Pearl, 2000; Spirtes et al, 2001; Heckerman, 1999).

*(i) Directed Acyclic Graph*: A directed acyclic graph (DAG) is a graph **G**($V$,$E$) where $V$ represents the vertices and $E$ represents an edge, such that each edge is a directed edge and no cycles are allowed. In this study, the vertices (*V*) are represented by the random variables (genes) and the edges (*E*) the dependencies between them.

A Bayesian network represents the joint distribution of a set of random variables, represented by the vertices (*V*) in a DAG *G* (Figure 7) as a product of conditional distributions. Consider a DAG $G : g_i, i = 1...v$, where the parents of vertex $g_i$ is $p(g_i)$; the joint distribution of $G$ is given by:

$$P(g_1,...,g_v) = \prod_{i=1}^{v} P(g_i / p(g_i)) \qquad (3)$$

*(ii) Markov Condition:* A necessary and sufficient condition for a probability distribution *P* to be Markov relative to a DAG G, is that every variable is independent of all its non-descendants conditional on its parents (Pearl, 2000).

*Remark*: There are two important assumptions in the definition of Bayesian networks. The first assumption is the *acyclic* nature of the graph, which assumes absence of cycles (feedbacks). The second assumption is that the joint distribution can be factored into a product of conditional dependencies given by (equation 3). In genetic networks, it is not uncommon to observe feedbacks (cycles) and the nonlinearly coupled nature of the networks might not lend themselves to be factored as a product of conditional dependencies. It should also be noted that the present study did not consider hidden variables, such as proteins that are an inherent part of the biological network. Thus the resulting network is an *approximation* of the actual biological network.

*(iii) Markov Equivalence:* (Pearl, 2000) defines two DAGs to be Markov equivalent if they have the same sets of $v$-structures whose tails are not connected by an arrow.

Markov equivalence addresses the important question as to whether the observed structure is *unique* or is it a part of an *equivalent class* of structures. A simple example of Markov equivalence is exhibited by the DAGs ($G_1$) and ($G_2$) where $G_1$: $g_1 \rightarrow g_2$ and $G_2$: $g_2 \rightarrow g_1$. The corresponding joint probability distributions if factored into conditional dependencies are given by: $p(g_1)\,p(g_2 / g_1)$ and $p(g_2)\,p(g_1 / g_2)$ respectively. By Baye's theorem each of the joint probability distributions is equal to $P(g_1, g_2)$. In the first case, an upstream gene $g_1$ affects the value of a downstream gene $g_2$, whereas the roles are reversed in the second case. Thus, information of the conditional dependencies fails to provide insight into the *direction* of regulation or *causality* of the network. Therefore, the equivalent class is represented as an undirected graph $g_1 - g_2$. For a DAG $g_1 \rightarrow g_2$, manipulating the expression of gene $g_2$ has no effect on $g_1$, while it is not true the other way around. It is important to reiterate the fact that we do not consider feedback in these models. Gene knockout studies (interventions) may be used to infer such structures Friedman et al. (2000). However, there may be instances where even interventions *may* fail to provide insight into network structure. Consider the networks of the form $G_1$: $g_1 \rightarrow g_2 \rightarrow g_3$ and $G_2$: $\{g_1 \rightarrow g_2, g_1 \rightarrow g_3\}$, where knocking out $g_1$ results in no expression of $g_2$ and $g_3$. Either of the structures is a possible explanation of the result. However, a prudent choice of gene knockout can be helpful in inferring network structure, Akutsu et al. (1998).

A more classic, and an important Markov equivalence class, is shown in Figure 7. For the chains Figures 7 (a and c) and the fork Figure 7b, the marginally dependent genes *A* and *C* become independent, conditioned on *B*, i.e. learning about *A* does not affect the state of *C*, given *B*. Alternatively, these DAGs are Markov equivalent and it is not possible to infer the direction of the arcs based on conditional probability distributions. The equivalence structure representing the DAGs in Figure 7 is given by ($A – B – C$). The DAG in Figure 7d is a $v$-structure, also known as the *collider* (Pearl, 2000). Unlike that of the forks and the chains, the equivalent class of collider consists only the collider, thus retains the direction of the arcs. It is important to note that the tails of the collider should be free and not connected by an arrow, as per the definition of Markov equivalence (Pearl, 2000). If the tails are connected, the possible directions of the tail are $A \to C$ and $C \to A$. The former results in a DAG with joint distribution $p(B/AC)p(A)p(C/A)$, while the latter has joint distribution $p(B/AC)p(C)p(A/C)$. Since $p(C)p(A/C) = p(A)p(C/A) = p(AC)$, the two structures are equivalent. Thus it is important that the tails are not connected by an arrow. Contrary to the chains and forks, in a collider, two marginally independent genes, *A* and *C*, become dependent conditioned on gene *B*. This result is known as *Berkson's paradox* or *explaining away effect* (Pearl, 2000). There are two possible causes that explain the state of *C*. The most plausible one explains away the other. Pearl, 2000 introduced the concept of *d-separation* and *pattern*, which emphasizes the fact that a given distribution has a unique minimal causal structure up to d-separation. Thus, it is crucial to determine the equivalent class of a network, as conclusions based on single realizations is incomplete. This is evident from the example shown in Figure 7, where a, b and c are equivalent structures and cannot be distinguished by probablisitic information alone. Such an equivalent class of networks is also represented as *partially directed acyclic graphs* or PDAGs. They contain both directed and undirected edges, hence the term PDAG. While the directed edges of the PDAGs are common to all the DAGs in the equivalent class, it is not true the other way around. However, it is important to understand that a directional edge of the form $X \rightarrow Y$ is more of an indication rather than evidence that $X$ is the cause of $Y$ (Pearl, 2000; Spirtes et al., 2000; Glymour and Cooper, 1999; Friedman et al 2000).

*Summary (Sec. 4): Given the above intricacies, supported by examples in Figure 7, it is clear that identifying pair-wise dependencies may be inadequate in determining the underlying network structure. This is attributed to the fact that some of these dependencies may break down conditioned on the other members of the network. It is also important to note that the equivalent class of networks (PDAGs) as conclusions based on (DAGs) is incomplete.*

### *4.1 Learning network structure with varying sample sizes*

Several techniques have been proposed in the past for learning the structure of a Bayesian network from a given observation. These include exhaustive enumeration, constraint-based approaches, and search-based techniques (Pearl, 2000; Spirtes et al., 2000; Heckerman, 1999, Glymour and Cooper, 1999). The number of possible DAGs increases super-exponentially with the number of nodes; hence exhaustive enumeration is not feasible for networks with nodes (> 5) Glymour and Cooper, (1999). In this study, constraint-based methods (Spirtes et al., 2000) along with search techniques such as hill climbing and Markov Chain Monte Carlo (MCMC) (Heckerman, 1999; Glymour and Cooper, 1999) were used to determine the network structure.

*4.1.1. Learning network structure - Constraint based technique*

The PC (Peter and Clark) algorithm by Spirtes et al. (2001) determines the underlying network structure from the conditional independence information. The method begins with all possible relationships between the genes and determines the network structure by systematically deleting the edges. The edge removal is based on conditional independence criteria. In this study, the Fisher's Z test was used to determine the statistical significance of the conditional dependencies. The maximum number of incoming edges to any node was fixed at three, the significance level at ($\alpha = 0.05$) and ($N_B = 1000$) bootstrap realizations were generated for each sample size. These bootstrap realizations represent perturbations of the given data (Friedman et al. 2002). The frequency of occurrence of a (directional) dependency of the form $X \rightarrow Y$ in the resulting PDAGs $G_p$ was determined with the expression:

$$P_{x \rightarrow y} = \frac{\text{Number of occurences of } X \rightarrow Y \text{ in } G_p(i), i = 1 \ldots N_B}{N_B} \qquad \ldots\ldots\ldots\ldots\ldots\ldots\ldots(4)$$

Similarly, one can determine $P_{x \leftarrow y}$. The frequency of occurrence of a dependency between nodes $X$ and $Y$ is given by $P_{XY} = P_{x \rightarrow y} + P_{x \leftarrow y} + P_{x - y}$. The dependencies generated by the constraint-based approach were ranked according to $P_{XY}$. The values of $P_{XY}$, $P_{x \rightarrow y}$ and $P_{x \leftarrow y}$ is shown in Table II. While the ranking of the dependencies is consistent with decreasing sample size, the power reduced considerably (Table II). The resulting network contained interaction between the genes *ALP*, *BSP*, *COLL* and *OCN*. The dependencies

*COLL – ALP; ALP – OCN; ALP – BSP* have power > 0.8 in bootstraps generated with sample sizes of 66 and 99 colonies, whereas a marked decrease is observed with sample size 33. From Fig. 8, it can be observed that neither *PDGFRa* nor *PTHrP* is a part of the network and may act independently of other genes (Madras et al., 2002). While the frequency of occurrence can be thought of as the statistical *power,* it is not clear what would be an acceptable value (Friedman et al., 2002). From Table II, it should be noted that the frequency of occurrence $COLL \rightarrow ALP$ is consistently greater than the frequency of occurrence of $ALP \rightarrow COLL$, similarly for $ALP \rightarrow OCN$ and $BSP \rightarrow ALP$. Thus it is highly likely that the edges are oriented accordingly. An approximate network structure from the colony data constructed with the constraint-based approach and the probability of occurrence of the edges in the PDAGs, Table II, is shown in Figure 8.

### *4.1.2 Learning network structure – Search based technique*

While the constraint-based approach determines the network structure from conditional dependencies, search algorithms along with a scoring criteria can be used as an effective alternative. Hill climbing, is a heuristic local search technique (Heckerman, 1999), that determines the network structure by adding, deleting or reversing the edges and evaluating the score with each change. The objective is to find the structure that maximizes the scoring function, i.e. maximize $P(S/D)$, where $S$ is the structure given the data $D$, but $P(S/D) = ( P(D/S)\ P(S) ) / P(D)$. It should be noted that the term $P(D)$ is a constant and can be filtered out. Thus maximizing log $P(S/D)$ is equivalent to maximizing log $P(D/S)$ + log $P(S)$. This is known as the Bayesian score of the structure $S$. In this study, we use the Bayesian score with Bayesian Dirichlet priors [Cooper and Herskovits, 1992; Heckerman, 1999]. However, such local search techniques such as hill climbing are susceptible to local optimum. As in Sec. 4.1.1, $N_B$ = 100 bootstrap realizations were generated for this analysis. The PDAGs of the DAGs (from the hill climbing procedure) were generated using the Chickering (Chickering, 1996) and the d-separation algorithms (Pearl, 2000). The maximum number of incoming edges was fixed as three for the d-separation criteria.

The steps in determining the dependencies are as follows:

- Determine the DAG $G(i)$ from the given data ($D$) using the hill-climbing approach
- Determine the PDAG $G_p(i)$ of the DAG $G(i)$

Repeat the above for $i = 1 \ldots N_B$, where $N_B$ represents the number of bootstrap realizations. Determine the frequency of occurrence of the edges using expression (equation 4).

The frequency of occurrence of the edges was determined with varying sample sizes (99, 66 and 33 colonies, with $N_B$ = 100) and subsequently ranked, Table III. The top three values corresponding to the edges $ALP - OCN$, $BSP - ALP$ and $ALP - COLL$ are shown in Figure 9. The frequency of occurrence of the edges $COLL - FGFR1$, $OCN$ – $PTH1R$, $OCN - BSP$ were similar (~0.5) across the different sample sizes and hence unclear. Unlike the case of 99 and 66 colonies, the frequency of occurrence of the edges $ALP - OCN$, $BSP - ALP$ and $ALP - COLL$ were below 0.8 in the case of 33 colonies. The PDAGs generated in the search and score method also did not provide information on the directionality of the edges. Finally, a Markov-chain Monte-Carlo (MCMC) implemented with Metroplis-Hasting algorithm [Metropolis et al., 1953; Hastings, 1970] was used to determine the network structure. The given data containing all the 99 colonies were used. Determining the convergence of the MCMC is a non-trivial issue and beyond the scope of the present report. In MCMC, the acceptance or rejection of the proposed random sample is determined by the acceptance ratio [Metropolis et al., 1953; Hastings, 1970]. A large value of the acceptance ratio implies that the proposed sample is close to the current state, whereas a small value implies proposals are distant from the current state. The acceptance ratio was plotted against the MCMC steps to determine the convergence (Figure 10). The burn-in was chosen as 500, i.e. the first 500 MCMC steps were assumed to be transients, hence discarded. The number of MCMC steps was fixed at 10,000. The DAGs generated in the last 1000 steps by the MCMC was chosen for subsequent analysis. The PDAGs of the DAGs obtained from the MCMC was generated using the Chickering (Chickering, 1996) and the d-separation algorithms (Pearl, 2000). The maximum number of incoming arcs was fixed at three for the d-separation criteria. The frequency of occurrence of the edges was determined using expression (4) and shown in Table IV. The approximate network structure of the genetic network is shown in Figure 11.

*Summary Sec.4.1: The network structure consisting of the undirected edges* $ALP - OCN$, $BSP - ALP$ *and* $ALP - COLL$ *was found to be robust to perturbations and reduced sample size. This was verified using constraint-based and search-based techniques. It is highly likely that these represent the set of robust dependencies and may form the backbone of the osteoblast differentiation program.*

**5. Discussion:**

Several techniques have been proposed in the past to model the network and the dependency between a given set of genes [Kauffman, 1993; Gardner et al., 2000; Shmulevich et al., 2002; Liang et al., 1998; Akutsu et al., 1998]. Each of these techniques works under certain implicit assumptions for the gene expression data. More importantly, they require explicit information regarding temporal gene expression and/or truth tables which might require gene knockout experiments. Data mining approaches can be used to obtain representation of the truth table, however a gene can be up-regulated in apoptosis (cell death) and cancer (cell proliferation) depending in the experimental paradigm. Thus, there are practical difficulties associated with these techniques. Nonlinear dynamical approaches such as those discussed in [Kauffman, 1993; Gardner et al., 2000] are highly involved. They have been used successfully to model interactions at transcriptional level in prokaryotes and hence represent true biological network, unlike the above which use measures of association. Such detailed models can form as a follow up to confirm the results obtained in this study.

The methods discussed here, infer the dependencies and network structure from gene expression across replicate clones without explicit temporal information. The data used in this study is in the form of a binary matrix consisting of nine genes across ninety-nine colonies during osteoblast differentiation, where each gene was assigned an ON or OFF state in a given colony. However, the methods can be extended to other types of data sets too. The objective was to determine the uncertainty, dependencies and structures that are robust to perturbations and reduction in the sample size of nine genes across ninety nine colonies in osteoblast differentiation. This was accomplished with entropy, linear correlation, mutual information and Bayesian networks. Perturbation of the data was accomplished by resampling with replacement (Sec. 2.2.1). The methods are generic and can be extended to gene expression data obtained in similar experimental paradigms. These include data sets obtained using tools such as microarrays [Friedman et al., 2000]. However, there are subtle issues associated with microarray analysis, such as: image segmentation, normalization, high dimensionality and lower number of replicates. Image segmentation and normalization form the preliminary steps and any error incurred at this stage is bound to propagate through subsequent analysis. Determining the optimal structure with a large-number of genes (dimension) and the small replicate arrays can be challenging if not impossible. This can be attributed to increasing local optimum with the number of genes. In the present study, any conclusion based on number of colonies lesser than 33 is likely to be inconclusive. However, microarrays can be used as a screening tool to subsequent clonal analysis of a small set of genes.

The ranked entropy profile of the nine genes in decreasing order was [*ALP, BSP, OCN, COLL, PDGFRa, FGFR1, PTHrP, PTH1R*] (Sec. 3.1) and did not change with perturbations and reduction in the sample size. The significant pair-wise dependencies obtained using linear correlation and Fisher's exact test ($\alpha$ = 0.05, with power ~ 0.8), across ($n_c$ = 33, 66 and 99) were: *(COLL–ALP), (ALP–BSP), (ALP–OCN),* and *(BSP–OCN).* Those obtained with mutual information (power ~ 0.8) were *(COLL–ALP), (COLL–OCN), (ALP–OCN), (ALP–BSP), (BSP–OCN)* and *(OCN – PTH1R).* As reported earlier, [Madras et al., 2002], the genes *PDGFRa* and *PTHrP* were independent of each other and the other genes. These were consistent across all the three measures of correlation. The mutual information was robust compared to the correlation measure in retaining the dependencies with perturbations and reduction in sample size, (Sec. 3). The pair-wise dependency analysis implicitly assumes each pair to be mutually exclusive and hence may be unhelpful in determining the network. It might not be surprising to note that certain dependencies may cease to exist when connected together as a part of the network. In order to capture the conditional dependencies we used the concepts of Bayesian networks (Sec. 4). The directed acyclic graphs (DAG) were determined using constraint-based and search-based techniques. The PC algorithm, which is a constraint-based technique, was used to determine the DAG and PDAG structures. Two search-based techniques, namely, hill-climbing and Markov-Chain Monte-Carlo (MCMC), were used to determine the DAGs. The PDAGs were generated from these DAGs using d-separation and Chickering's algorithm separately. Inferences based on DAG structures are incomplete, as conditional probabilistic distributions cannot distinguish equivalent structures (Sec. 4). The PDAGs represent structures (possibly causal) common to all the DAGs of a given equivalent class and hence its choice. To rule out the possibility that the given PDAG is by chance, the frequency of occurrence of the edges was determined from multiple realizations of the PDAGs. The structure *(COLL–ALP), (BSP–ALP)* and *(ALP–OCN)* was preserved across the PDAGs generated by the different techniques. The genes *PDGFRa* and *PTHrP* were independent of each other and the other members of the network, hence not a part of the final structure. This structure had a (power > 0.8) with decreasing sample size (99 and 66 colonies) and was robust to perturbation of the data. These structures were retained even when the number of boostrap realizations were changed and with exhaustive enumeration with five nodes comprising of (*COLL, ALP, BSP, OCN* and *PTH1R*), results not included. Therefore the structure *(COLL–ALP), (BSP–ALP)* and *(ALP–OCN),* might play a pivotal role in the osteoblast differentiation program. (Figures 8, 9 and 11). While the frequency of occurrence of the arcs for the edges *(COLL–ALP), (BSP–ALP)* and *(ALP–OCN)* were considerably high even in the case of 33 colonies, they failed to statistically significant (< 0.8). Thus we suspect analysis with number of colonies lesser than 33, would be inconclusive.

*Biological Issues:* Of the eight genes included in the analysis, the four identified above (*COLL, ALP, BSP* and *OCN*) are characteristic of osteoblasts, either encoding enzymes or matrix components integral to differentiated osteoblast function. Further, many of the aspects of the PDAGs obtained by the different approaches correlated with experimental studies (Stein and Lian, 1993; Aubin et al., 1999; Madras et al., 2002). The constraint-based technique indicated *COLL*®*ALP* to be more frequent than *ALP*®*COLL* supporting the hypothesis presented by Madras et al. (2002) that *COLL* is expressed before *ALP*. Similarly, *BSP*®*ALP* occurred more frequently than *ALP*®*BSP*. ALP may not be the last of the genes to be expressed during differentiation as *ALP*®*OCN* was more frequent than *OCN*®*ALP* (Table II and Figure 8), consistent with experimental studies (Stein and Lian, 1993; Malaval et al., 1999). That *ALP* was correlated with each of the other three genes (*COLL, BSP* and *OCN*) suggests that it is essential to osteoblast function (reviewed by Aubin et al., 1999), a conclusion supported by the observation that all bones display mineralization defects in *ALP* knockout mice (Fedde et al., 1999; Wennberg et al., 2000). It should be noted that the *BSP – OCN* and *COLL – OCN* dependencies came through some but not all of the measures, and may reflect the fact that all three genes contain Cbfa1 binding sites in their promoters (Ducy et al., 1997) and thus, appear to share common transcriptional regulatory mechanisms. The search-based approach by MCMC suggested that *COLL-FGFR1* and *OCN-PTH1R* may be components of the network (Table IV and Figure 11). Both FGF and PTH drive osteoblastogenesis, but in the network described here, their receptors appear to behave marginally independent and thus may utilize parallel pathways or receptors, consistent with the idea of exchangeability (Madras et al., 2002). On the other hand, as subpopulations of osteoblasts with distinct properties have been shown to exist (Liu et al., 1997; Candeliere et al., 2001), variation in the activity of FGF and PTH signaling pathways may contribute to osteoblast heterogeneity represented in the array of colonies analyzed here.

*Causality Issues*: The PDAGs were generated from the DAGs to infer possible causal structures common to the equivalent class. Causality is an indicator rather than a confirmation of temporal precedence of gene expression. Traditional time course experiments on a population of cells can be used to determine a temporal sequence of events (Stein and Lian, 1993). These in turn can be incorporated to determine the causal structure of the network obtained by the Bayesian approach. In this study, the variation in rates of differentiation between colonies incorporated a stochastic component using a single time point that may have a significant impact on the results.

*Reliability Issues*: From our analysis, it is clear that the genes *PTHrP* and *PDGFRa* are likely to behave independently of the other members of the network and from each other, consistent with the findings of Madras et al. 2002. This is confirmed by the pair-wise dependency study and the network structure. Therefore it is possible that these may not be required for osteoblast differentiation under the in vitro conditions utilized or may participate in parallel paths leading to the end phenotype. Determining the serial and parallel paths in the network can be useful in determining the network reliability. Tractable analytical expressions can be determined if the failure of the components in the networks is assumed to be independent of one another. For a serial network with dependent components (positively correlated), expression for failure time is conservative compared to the independent assumption, whereas it is anti-conservative for parallel networks. Under the assumption of a coherent network, consisting of *k* independent components, a bound can be achieved on the network reliability (Barlow and Proschan, 1981). The more parallel paths that may exist, the more reliable the network is to perturbations. This in turn brings up important questions as to whether some lineages are more reliable than others due to the very nature of their network. The reliability analysis also provides a way to quantify and distinguish across the various experimental conditions. Thus determining the network reliability assigns a quantitative value to the changing network structure.

*Nonlinear Dynamical Perspective*: While Bayesian networks is a powerful technique, it works under certain assumptions such as: acyclic nature of the network and representing the joint probability distribution as the product of conditional probability distributions. Feedback loops are ubiquitous in biological systems and their control can be explained only by nonlinear systems theory. A nonlinear dynamical perspective of the above problem would involve a nonlinearly coupled differential equation system, representing the interactions between the genes. The evolution and temporal behavior of a *k* gene network would be represented by a *k* dimensional state space (phase space). The class of dissipative dynamical systems, which most real world systems are, settles down on an *attractor* after an initial transient behavior. Such systems can exhibit a wide range of behaviors such as limit cycle (periodic), torus (quasiperiodic) and chaotic behavior. In the present experimental paradigm, each gene in the network can be represented by a variable. Thus, the interaction of *k* genes can be represented by a coupled nonlinear system consisting of *k* states, given by a set of differential equations. The evolution of such a system settles down on an attractor after an initial transient behavior. This attractor represents the end phenotype. The heterogeneity observed in the expression across the colonies can be attributed to *sampling* at the different points in the attractor, as the state of a gene here is represented by a continuum of points and gives rise to infinite possibilities. Strange attractors are hallmarks of chaotic systems, which show sensitivity to initial conditions. Such sensitivity can be induced by minor variations in external stimuli such as serum and hormonal cocktail concentrations, whose values are on the real plane. Theoretically, an infinite choice of the concentrations can give result in the same end phenotype (attractor). Thus another possible explanation for the observed heterogeneity between the colonies may be due to sampling and quantification of the state as being ON or OFF. As with Bayesian networks, a nonlinear dynamical approach also has its limitations. Stationarity issues and high-dimensionality of the system may limit its application to the present paradigm.

**Acknowledgements:**

We would also like to thank Kevin Murphy and Philip Leray for making available their Matlab routines, and the Arkansas Cancer Research Center (ACRC) differentiation group for useful discussions. This research was supported in part by funds provided to the UAMS Microarray Facility through Act 1, The Arkansas Tobacco Settlement Proceeds Act of 2000, by NIH Grant #P20 RR-16460 from the BRIN Program of the National Center for Research Resources and by an NIH grant to CAP (AG20941).

**Figure Legends:**

**Figure 1.** A block diagram showing the induction of noise at the various stages in gene expression data, where $\in_t$ and $h_t$ represent the dynamical and observational noise; $f$ and $g$ are nonlinear functions and $y_t$, the observed gene expression value.

**Figure 2.** The mean and the standard deviation of the entropy (self-information) of the eight genes, obtained by resampling 99 (top), 66 (middle) and 33 (bottom) colonies.

**Figure 3.** The mean and the standard deviation of the mutual information ($I$) obtained for the 28 pairs of genes, along with those obtained on the random shuffled surrogates (noise floor) is shown in Figure 3a (top). Those obtained using the correlation ($r$) measure along with the random shuffled surrogates (noise floor) is shown in Figure 3b (bottom). The value of the mutual information derived from the normal approximation, $I = -0.5 \log_2(1-r^2)$, is shown in Figure 3a for reference. 100 bootstrap samples with ninety-nine colonies each were used.

**Figure 4.** The mean and the standard deviation of the mutual information ($I$) obtained for the 28 pairs of genes, along with those obtained on the random shuffled surrogates (noise floor) is shown in Figure 4a (top). Those obtained using the correlation ($r$) measure along with the random shuffled surrogates (noise floor) is shown in Figure 4b (bottom). The value of the mutual information derived from the normal approximation, $I = -0.5 \log_2(1-r^2)$, is shown in Figure 4a for reference. 100 bootstrap samples with sixty-six colonies each were used.

**Figure 5.** The mean and the standard deviation of the mutual information (*I*) obtained for the 28 pairs of genes, along with those obtained on the random shuffled surrogates (noise floor) is shown in Figure 5a (top). Those obtained using the correlation ($r$) measure along with the random shuffled surrogates (noise floor) is shown in Figure 5b (bottom). The value of the mutual information derived from the normal approximation, $I = -0.5 \log_2(1-r^2)$, is shown in Figure 5a for reference. 100 bootstrap samples with thirty-three colonies each were used.

**Figure 6.** Power estimated using the linear correlation, Fisher's exact test and mutual information for each of the 28 pairs is shown in a, b and c respectively. The (solid line and the open circle) represents resampling with 99 colonies, and those obtained using 66 colonies and 33 colonies are represented by the (dotted lines) and (bold dashed lines) respectively. 100 bootstrap samples were generated for (33, 66 and 99 colonies).

**Figure 7.** The DAGs shown (a, b and c) form a Markov equivalent class of the form (*A* – *B* – *C*). The equivalence class can be generated by applying the Bayes theorem to the conditional dependencies shown in bold letters. The DAG (bottom) is an example of *v*-structure; unlike (a, b, c) it is unique.

**Figure 8.** Approximate structure (PDAG) constructed from the frequency of occurrence, Table II, using the constraint based approach. The arrows denote possible direction of edges. The vertices (C, A, B and O) represent the genes COLL, ALP, BSP and OCN respectively.

**Figure 9.** Approximate structure (PDAG) corresponding to the frequency of occurrence, Table III. The vertices (C, A, B and O) represent the genes COLL, ALP, BSP and OCN respectively.

**Figure 10.** Convergence of the acceptance ratio with MCMC steps.

**Figure 11.** Approximate structure (PDAG) constructed from the frequency of occurrence, Table IV, using Markov-Chain Monte-Carlo (MCMC). The PDAGs were determined using the Chickering algorithm and the d-separation criteria (Table IV). The vertices (C, A, B and O) represent the genes COLL, ALP, BSP and OCN respectively.

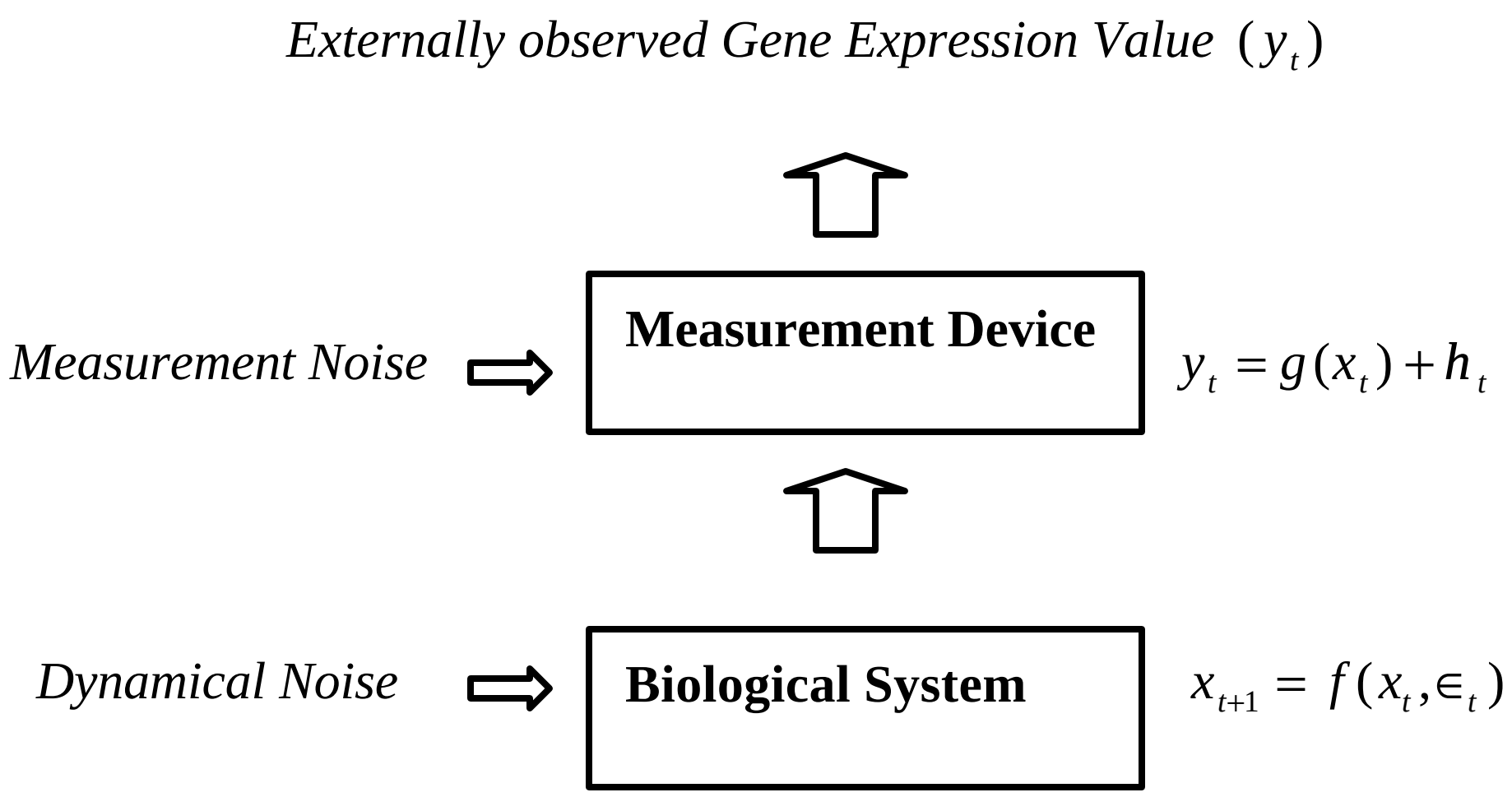
Externally observed Gene Expression Value $(y_t)$
Measurement Noise
Measurement Device
$y_t = g(x_t) + h_t$
Dynamical Noise
Biological System
$x_{t+1} = f(x_t, \in_t)$

Figure 1

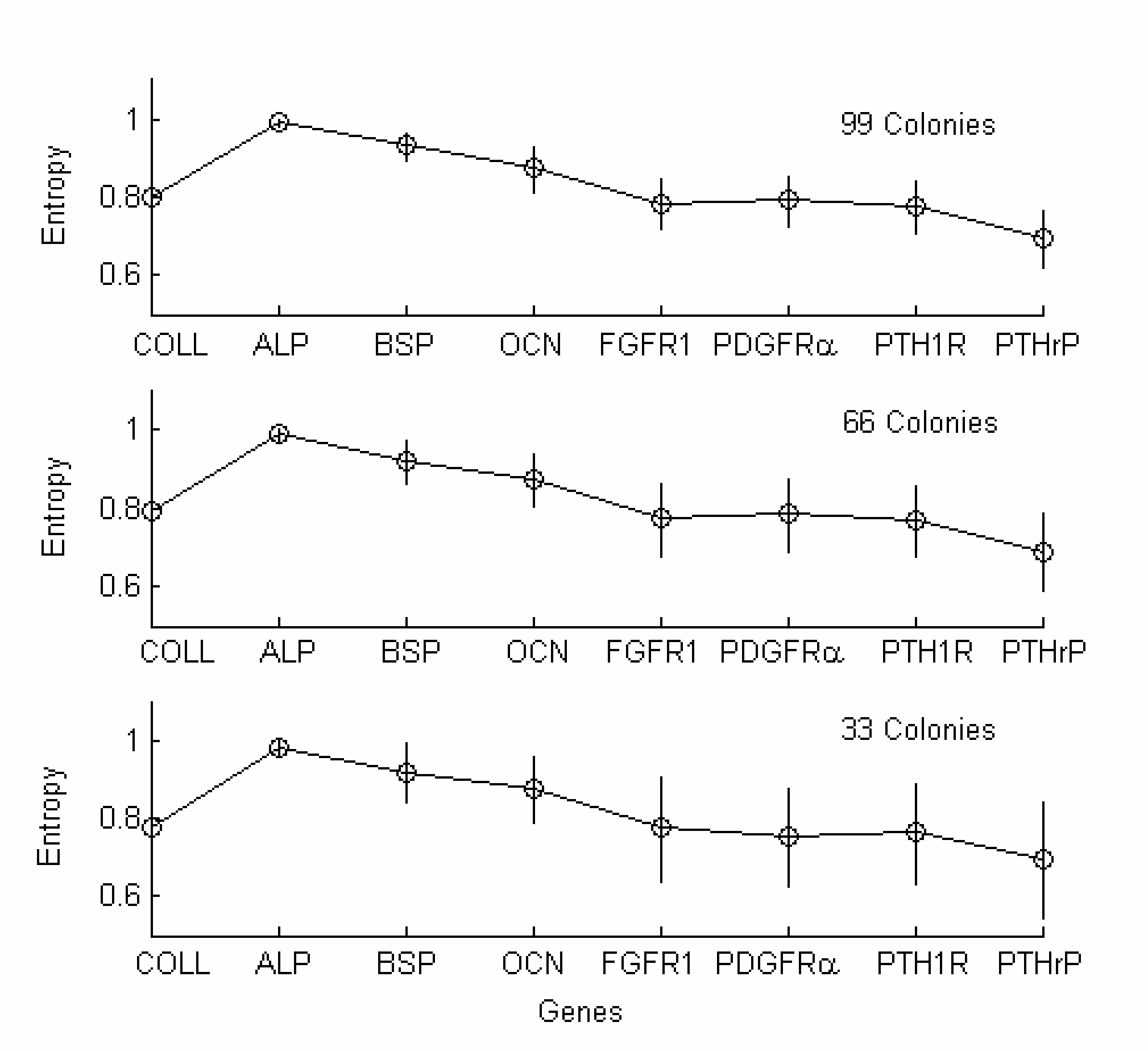

Figure 2

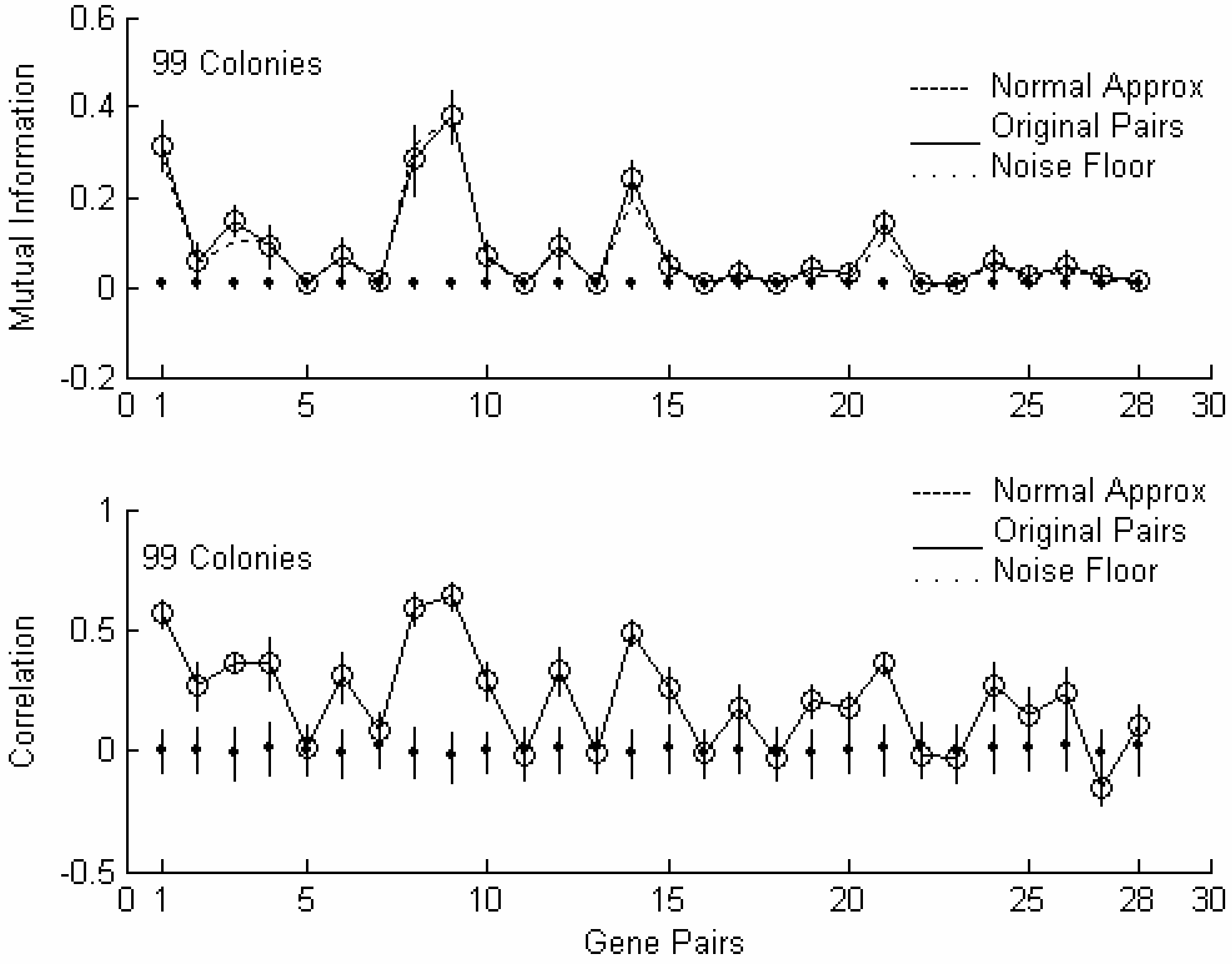
99 Colonies
Normal Approx
Original Pairs
Noise Floor
Mutual Information
0.6
0.4
0.2
0
-0.2
0 1 5 10 15 20 25 28 30
99 Colonies
Normal Approx
Original Pairs
Noise Floor
Correlation
1
0.5
0
-0.5
0 1 5 10 15 20 25 28 30
Gene Pairs

Figure 3

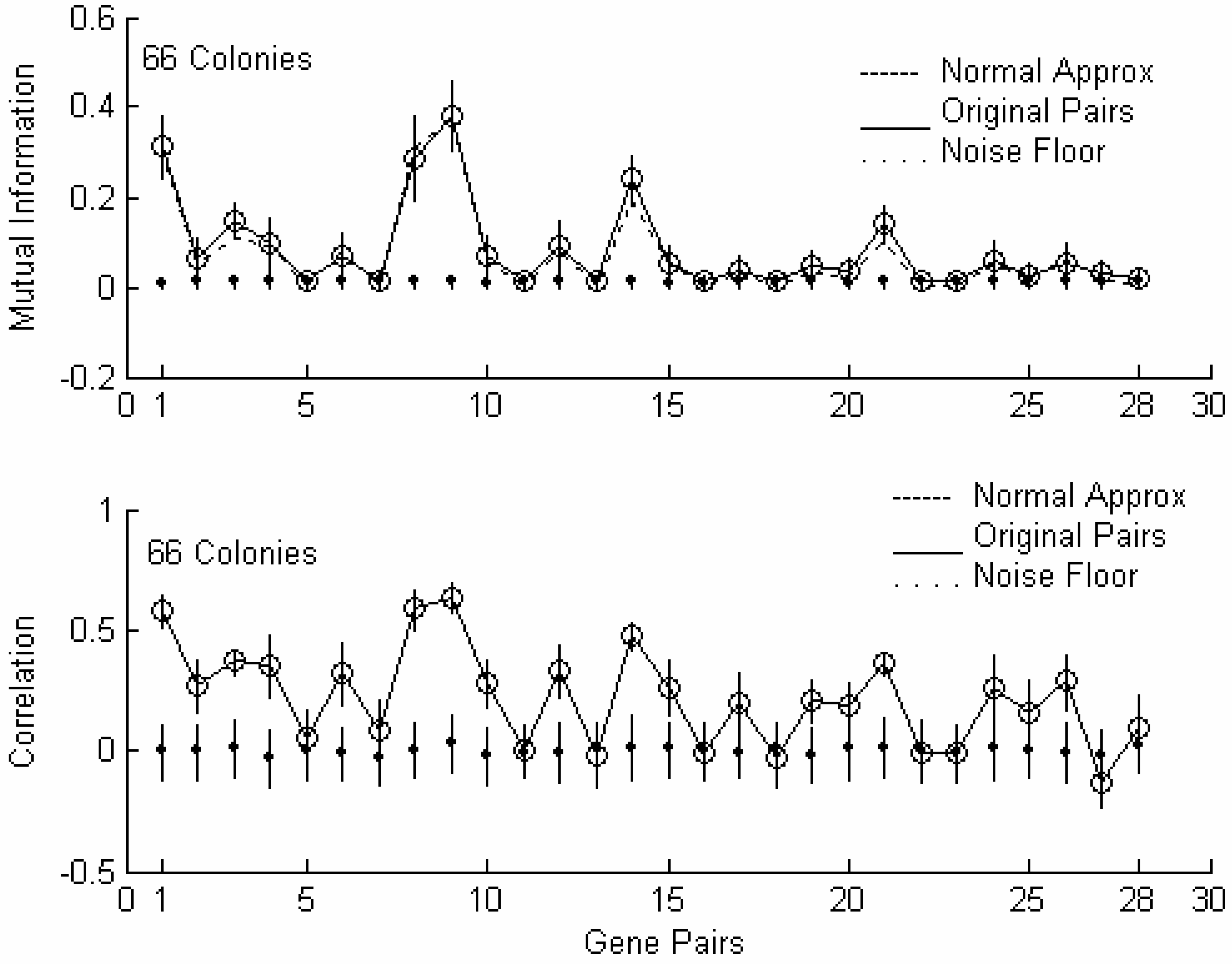

66 Colonies
Normal Approx
Original Pairs
Noise Floor
Mutual Information
0.6
0.4
0.2
0
-0.2
0 1 5 10 15 20 25 28 30
66 Colonies
Normal Approx
Original Pairs
Noise Floor
Correlation
1
0.5
0
-0.5
0 1 5 10 15 20 25 28 30
Gene Pairs

Figure 4

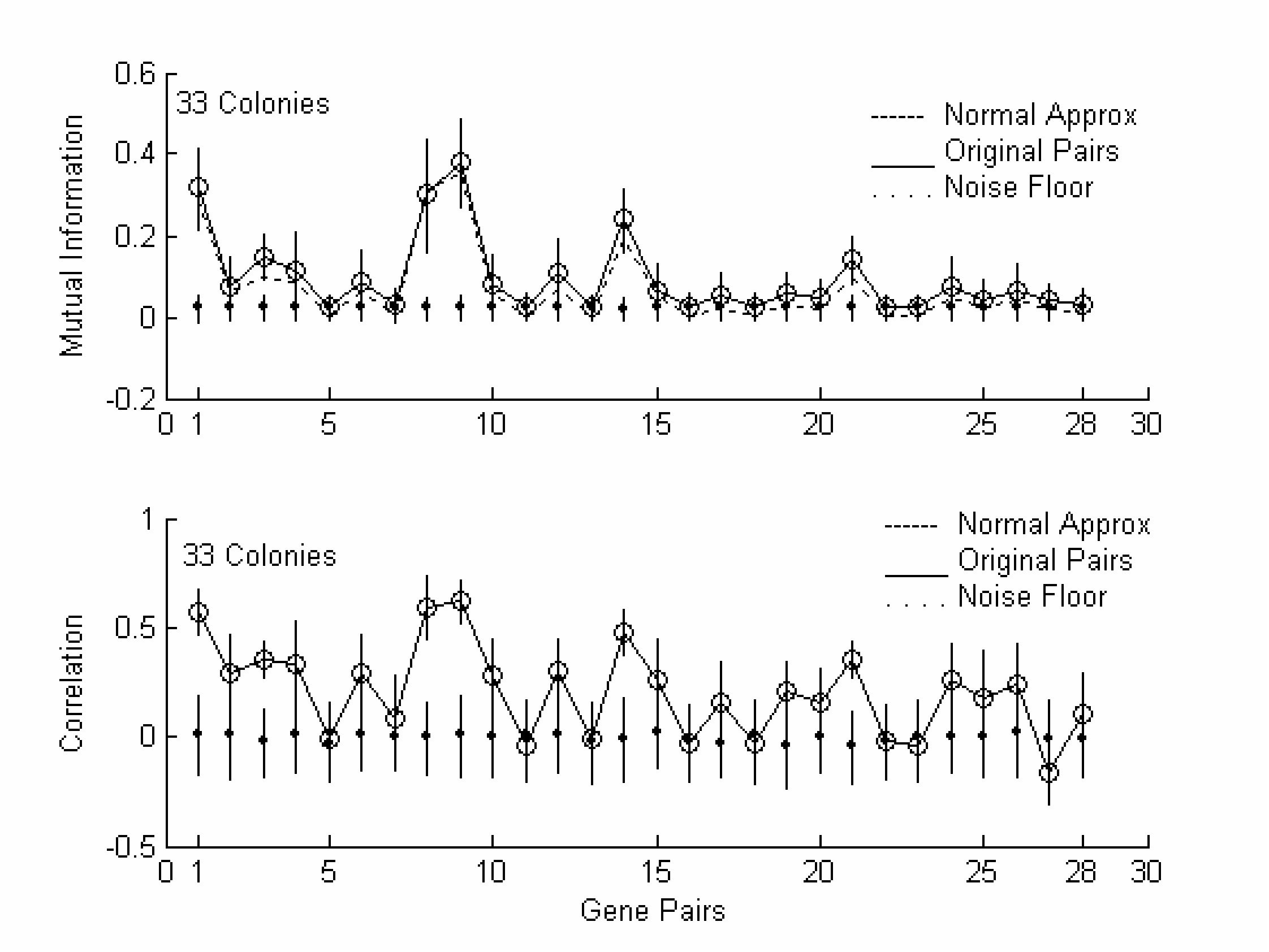

Figure 5

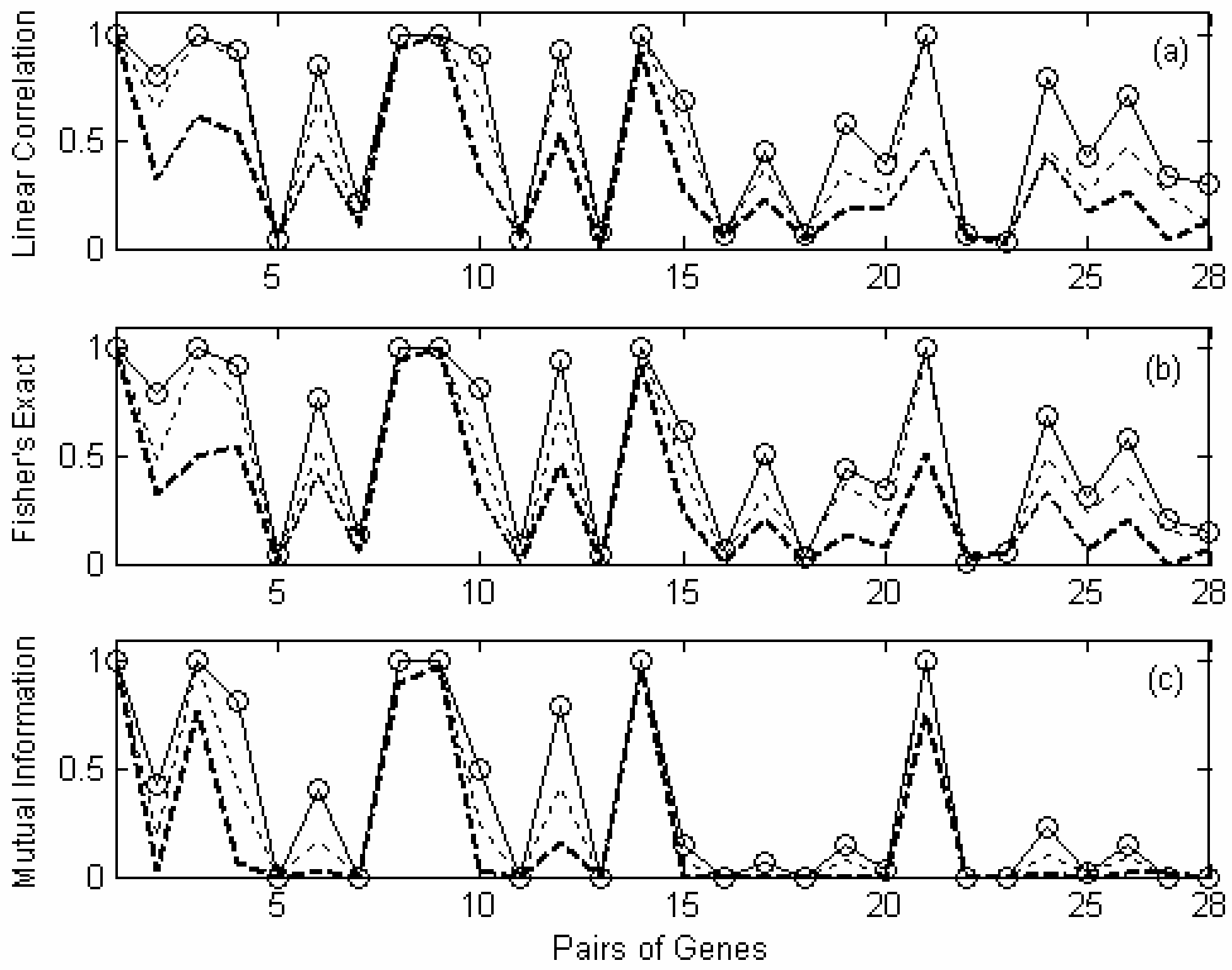

Figure 6

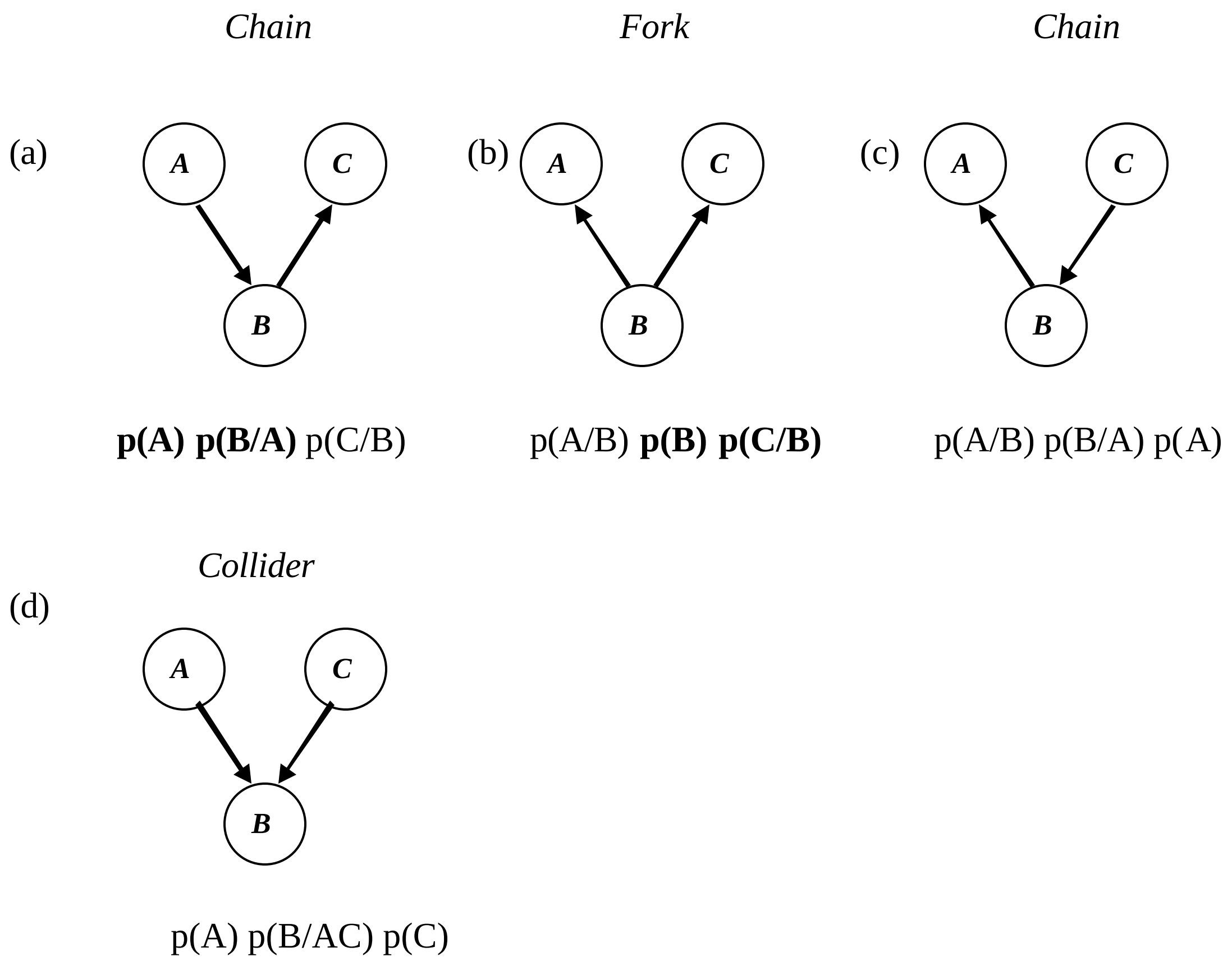

Figure 7

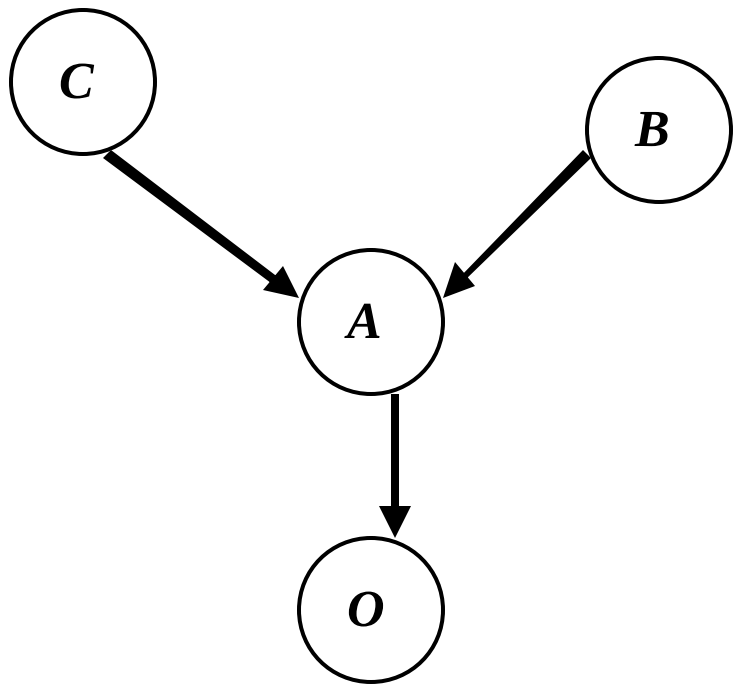

Figure 8

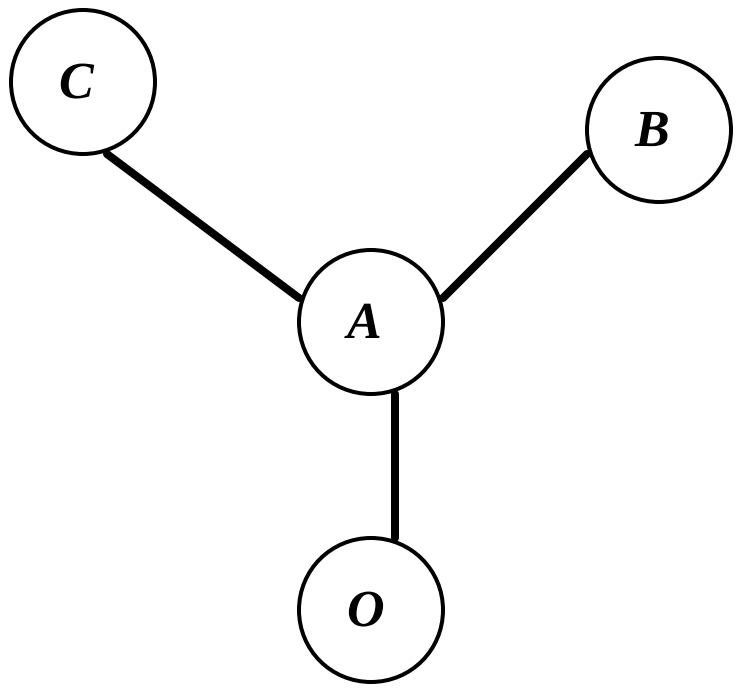

Figure 9

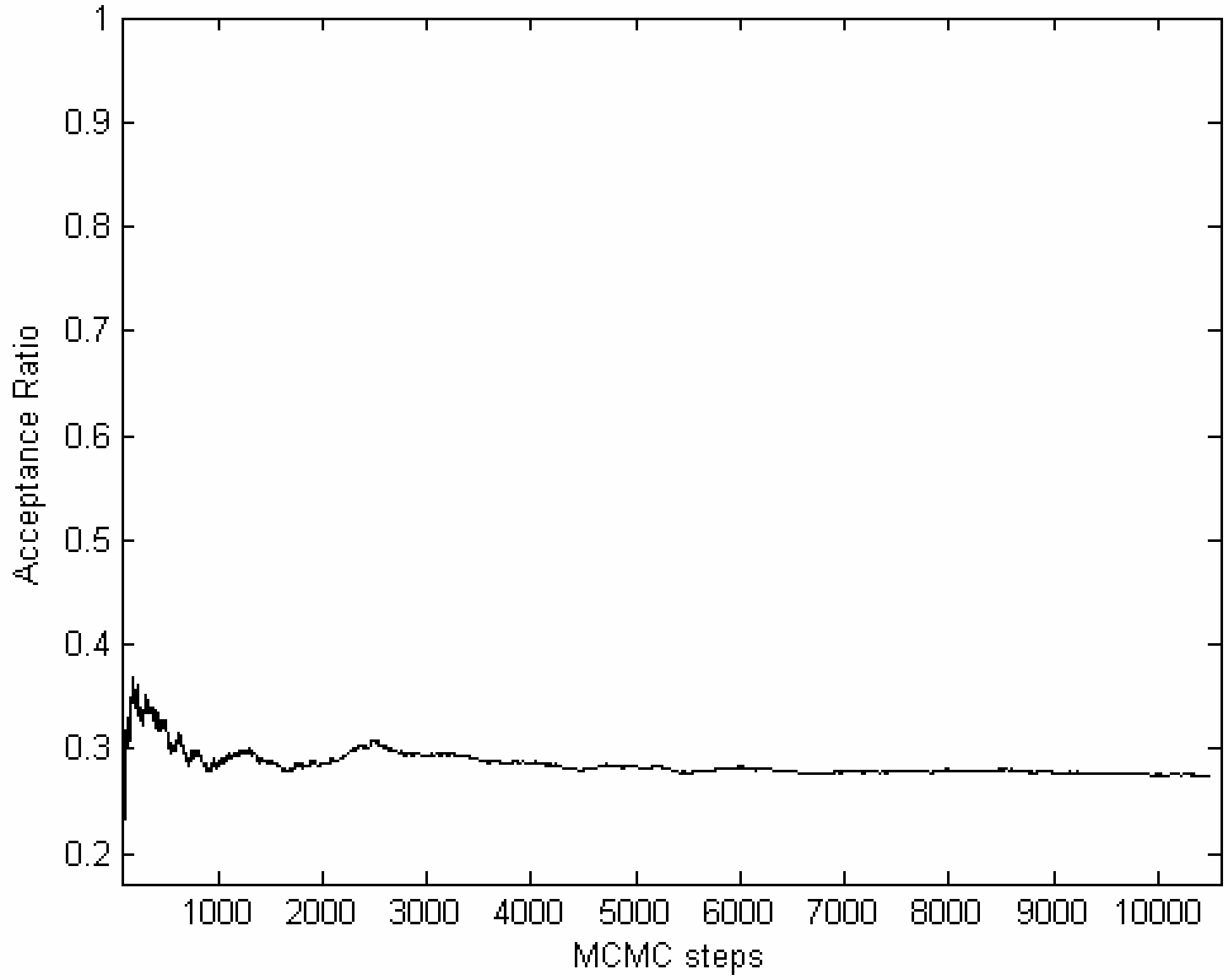
Acceptance Ratio
1
0.9
0.8
0.7
0.6
0.5
0.4
0.3
0.2
1000 2000 3000 4000 5000 6000 7000 8000 9000 10000
MCMC steps

Figure 10

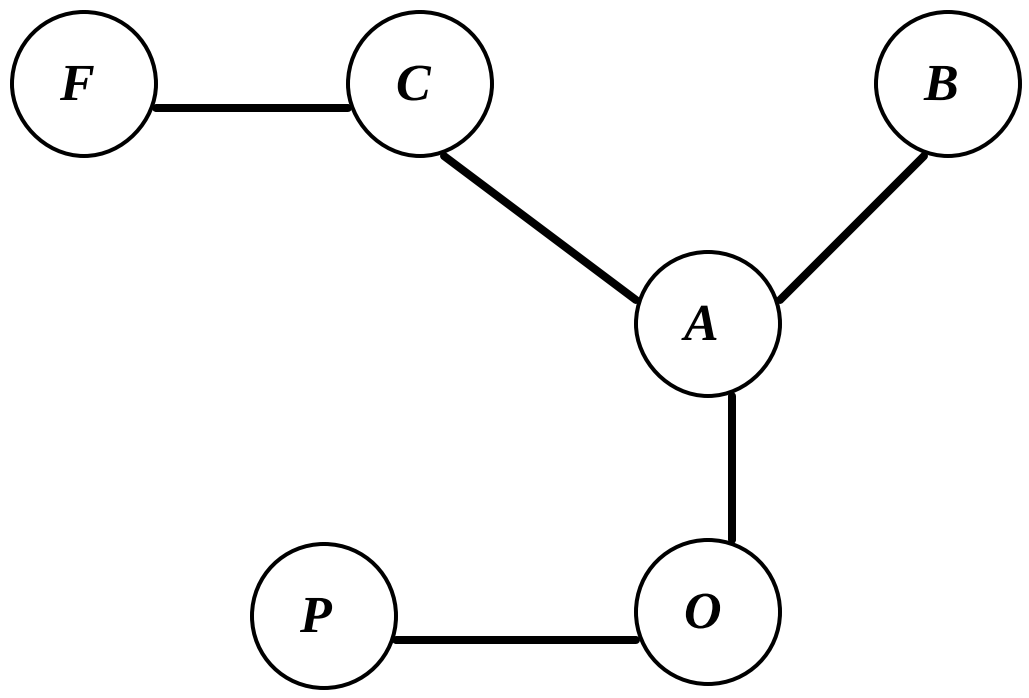

Figure 11

**Table I**

**Dependencies that had a (power ~ 0.8, a = 0.05) with ($N_B$ = 100) bootstrap realization using the linear correlation, mutual information and Fisher's exact test for the various colony sizes ($n_c$ = 99, 66, 33, 22 and 11). The genes *COLL*, *ALP*, *BSP*, *OCN*, *FGFR1*, *PTH1R*, *PDGFa* are represented by C, A, B O, F, P and D in the table.**

| Linear Correlation | | | | | Mutual Information | | | | | Fisher's Exact Test | | | | |
|---|---|---|---|---|---|---|---|---|---|---|---|---|---|---|
| 99 | 66 | 33 | 22 | 11 | 99 | 66 | 33 | 22 | 11 | 99 | 66 | 33 | 22 | 11 |
| **C-A** | **C-A** | **C-A** | **C-A** | - | **C-A** | **C-A** | **C-A** | **C-A** | - | **C-A** | **C-A** | **C-A** | **C-A** | - |
| **C-B** | - | - | - | - | - | - | - | - | - | **C-B** | - | - | - | - |
| **C-O** | **C-O** | - | - | - | **C-O** | **C-O** | **C-O** | - | - | **C-O** | **C-O** | - | - | - |
| **C-F** | **C-F** | - | - | - | **C-F** | - | - | - | - | **C-F** | **C-F** | - | - | - |
| **C-P** | - | - | - | - | - | - | - | - | - | **C-P** | - | - | - | - |
| **A-B** | **A-B** | **A-B** | **A-B** | - | **A-B** | **A-B** | **A-B** | - | - | **A-B** | **A-B** | **A-B** | **A-B** | - |
| **A-O** | **A-O** | **A-O** | **A-O** | - | **A-O** | **A-O** | **A-O** | **A-O** | - | **A-O** | **A-O** | **A-O** | **A-O** | - |
| **A-F** | - | - | - | - | - | - | - | - | - | **A-F** | - | - | - | - |
| **A-P** | **A-P** | - | - | - | **A-P** | - | - | - | - | **A-P** | **A-P** | - | - | - |
| **B-O** | **B-O** | **B-O** | - | - | **B-O** | **B-O** | **B-O** | **B-O** | - | **B-O** | **B-O** | **B-O** | - | - |
| **O-P** | **O-P** | - | - | - | **O-P** | **O-P** | **O-P** | - | - | **O-P** | **O-P** | - | - | - |
| **F-P** | - | - | - | - | - | - | - | - | - | - | - | - | - | - |

**Table II**

**Frequency of occurrence of the edges from PDAGs generated using constraint based approach. The vertices (C, A, B and O) represent the genes COLL, ALP, BSP and OCN respectively.**

| 99 colonies | | | | | | |
|---|---|---|---|---|---|---|
| | $XY$ | $P_{XY}$ | $X \rightarrow Y$ | $P_{X \rightarrow Y}$ | $X \leftarrow Y$ | $P_{X \leftarrow Y}$ |
| 1 | $CA$ | 1.00 | $C \rightarrow A$ | 0.32 | $C \leftarrow A$ | 0.09 |
| 2 | $AO$ | 0.99 | $A \rightarrow O$ | 0.51 | $A \leftarrow O$ | 0.03 |
| 3 | $BA$ | 0.95 | $B \rightarrow A$ | 0.29 | $B \leftarrow A$ | 0.16 |
| **66 colonies** | | | | | | |
| | $XY$ | $P_{XY}$ | $X \rightarrow Y$ | $P_{X \rightarrow Y}$ | $X \leftarrow Y$ | $P_{X \leftarrow Y}$ |
| 1 | $CA$ | 0.97 | $C \rightarrow A$ | 0.37 | $C \leftarrow A$ | 0.11 |
| 2 | $AO$ | 0.96 | $A \rightarrow O$ | 0.47 | $A \leftarrow O$ | 0.02 |
| 3 | $BA$ | 0.83 | $B \rightarrow A$ | 0.32 | $B \leftarrow A$ | 0.09 |
| **33 colonies** | | | | | | |
| | $XY$ | $P_{XY}$ | $X \rightarrow Y$ | $P_{X \rightarrow Y}$ | $X \leftarrow Y$ | $P_{X \leftarrow Y}$ |
| 1 | $AO$ | 0.61 | $A \rightarrow O$ | 0.14 | $A \leftarrow O$ | 0.10 |
| 2 | $CA$ | 0.55 | $C \rightarrow A$ | 0.27 | $C \leftarrow A$ | 0.05 |
| 3 | $BA$ | 0.46 | $B \rightarrow A$ | 0.22 | $B \leftarrow A$ | 0.04 |

**Table III**

**Frequency of occurrence of the edges in PDAGs obtained by the D-separation and the Chickering algorithm, on DAGs generated by Hill Climbing. The vertices (C, A, B and O) represent the genes COLL, ALP, BSP and OCN respectively.**

| **99 Colonies** | | | |
|---|---|---|---|
| | *XY* | $P_{XY}$ (D-Separation) | $P_{XY}$ (Chickering) |
| 1 | *CA* | 0.98 | 0.99 |
| 2 | *AO* | 0.83 | 0.83 |
| 3 | *BA* | 0.83 | 0.83 |
| **66 Colonies** | | | |
| 1 | *CA* | 0.98 | 0.98 |
| 2 | *AO* | 0.90 | 0.83 |
| 3 | *BA* | 0.75 | 0.75 |
| **33 Colonies** | | | |
| 1 | *CA* | 0.75 | 0.75 |
| 2 | *AO* | 0.75 | 0.76 |
| 3 | *BA* | 0.66 | 0.66 |

**Table IV**

**Frequency of occurrence of the edges in PDAGs obtained by the D-separation and the Chickering algorithm, on DAGs generated by Monte-Carlo Markov Chain (MCMC). The vertices (C, A, B, O, F and P) represent the genes COLL, ALP, BSP, OCN, FGFR1 and PTH1R respectively.**

| | $XY$ | $P_{XY}$ (Chickering) | $P_{XY}$ (D-Separation) |
|---|---|---|---|
| 1 | $AO$ | 1.0 | 1.0 |
| 2 | $CA$ | 1.0 | 1.0 |
| 3 | $BA$ | 1.0 | 1.0 |
| 4 | $OP$ | 0.94 | 0.94 |
| 5 | $CF$ | 0.85 | 0.85 |
| 6 | $BO$ | 0.75 | 0.75 |